\pdfoutput=1

\documentclass[11pt,preprint]{aastex}
\slugcomment{Accepted to PASP}

\def\kms{\ifmmode{\rm km\thinspace s^{-1}}\else km\thinspace s$^{-1}$\fi}
\def\ms{\ifmmode{\rm m\thinspace s^{-1}}\else m\thinspace s$^{-1}$\fi}

\newcommand{\msun}{\ensuremath{M_\sun}}

\newcommand{\rjup}{\ensuremath{R_{\rm J}}}
\newcommand{\mjup}{\ensuremath{M_{\rm J}}}
\newcommand{\teff}{\ensuremath{T_{\rm eff}}}

\newcommand{\re}{\ensuremath{R_{\rm E}}}

\shortauthors{Howell et al.}
\shorttitle{K2 Mission}

\begin{document}

\title{The K2 Mission: Characterization and Early Results}

\author{
Steve~B.~Howell,\altaffilmark{1} 
Charlie Sobeck,\altaffilmark{1} 
Michael Haas,\altaffilmark{1} 
Martin Still,\altaffilmark{1,2} 
Thomas Barclay,\altaffilmark{1,2} \\ 
Fergal Mullally,\altaffilmark{1,3} 
John Troeltzsch,\altaffilmark{4} 
Suzanne Aigrain,\altaffilmark{5} 
Stephen T. Bryson,\altaffilmark{1} \\
Doug Caldwell,\altaffilmark{1,3} 
William~J. Chaplin,\altaffilmark{6,11} 
William~D. Cochran,\altaffilmark{7} 
Daniel Huber,\altaffilmark{1,3} \\
Geoffrey~W. Marcy,\altaffilmark{8} 
Andrea Miglio, \altaffilmark{6,11} 
Joan~R. Najita,\altaffilmark{9} 
Marcie Smith,\altaffilmark{1} 
J.D. Twicken,\altaffilmark{1,3} \\
Jonathan~J. Fortney\altaffilmark{10}
}

\altaffiltext{1}{NASA Ames Research Center, Moffett Field, CA 94035, USA}
\altaffiltext{2}{Bay Area Environmental Research Inst., 560 Third St., West Sonoma, CA 95476, USA}
\altaffiltext{3}{SETI Institute, 189 Bernardo Avenue, Suite 100, Mountain View, CA 94043, USA}
\altaffiltext{4}{Ball Aerospace and Technology Corp., P.O. Box 1062, Boulder, CO 80306, USA} 
\altaffiltext{5}{Sub-department of Astrophysics, Department of Physics, University of Oxford, Oxford OX1 3RH}
\altaffiltext{6}{School of Physics and Astronomy, University of Birmingham, Edgbaston, Birmingham, B15 2TT,
UK} 
\altaffiltext{7}{McDonald Observatory \& Department of Astronomy, The University of Texas at Austin, Austin,
TX 78712, USA} 
\altaffiltext{8}{University of California, Berkeley, CA 94720, USA} 
\altaffiltext{9}{National Optical Astronomy Observatory, 950 N. Cherry Avenue, Tucson, AZ 85719, USA}
\altaffiltext{10}{Department of Astronomy and Astrophysics, University of California, Santa Cruz, CA 95064,
USA}
\altaffiltext{11}{Stellar Astrophysics Centre (SAC), Department of Physics and Astronomy, Aarhus
University, Ny Munkegade 120, DK-8000 Aarhus C, Denmark}


\begin{abstract} 

The K2 mission will make use of the {\it{Kepler}} spacecraft and its assets to 
expand upon {\it{Kepler's}} groundbreaking
discoveries in the fields of exoplanets and astrophysics through new and exciting observations. 
K2 will use an innovative way of operating the spacecraft to observe target fields along the ecliptic
for the next 2-3 years. Early science commissioning observations have shown 
an estimated photometric precision near 400 ppm in a single 30 minute observation, and
a 6-hour photometric precision of
80 ppm (both at V=12). The K2 mission offers long-term, simultaneous
optical observation of thousands of objects at a precision
far better than is achievable from ground-based telescopes.
Ecliptic fields will be observed for approximately 75-days enabling
a unique exoplanet survey which fills the gaps in
duration and sensitivity between the {\it{Kepler}} and
TESS missions, and offers pre-launch exoplanet
target identification for JWST transit spectroscopy.
Astrophysics observations with K2 will include studies of young open clusters, bright stars, 
galaxies, supernovae, and asteroseismology.
 \end{abstract}

\section{Introduction}

The NASA {\it{Kepler}}  mission was launched in 2009 and collected wide-field photometric observations of a 
single field of view located in the constellations of Cygnus and Lyra (Borucki, et al., 2010). 
The data obtained by {\it{Kepler}}
has revolutionized the study of exoplanets and astrophysics by providing high-precision, 
high-cadence, continuous light curves of tens of thousands of stars.
The loss of two reaction wheels on the {\it{Kepler}} spacecraft has ended the primary mission data
collection and the project is currently analyzing the final year of observations. The {\it {Kepler}} project has thus proposed the
K2 mission to NASA via the 2014 Senior Review process. K2 is the name chosen for this new mission concept to
allow a distinction between it and the nominal {\it {Kepler}} mission. K2 can be thought of as being a 2-wheel {\it {Kepler}},
the second {\it {Kepler}} mission, or as an analogy to the enigmatic and challenging mountain of the same name.
A decision on funding K2 will be reached in late spring 2014 and 
if positive, K2 will begin official operations near 1 June 2014, making observations for approximately 
the next 2 years along the lines discussed in this paper.

The {\it{Kepler}} spacecraft, with its 0.95-m Schmidt telescope and 110 sq. degree field of view
imager (4 arc-second pixels), is in a heliocentric orbit currently about 0.5 AU from the Earth.
Like {\it{Kepler}}, the K2 mission is founded on the proven
value of long-baseline, high-cadence, high-precision
photometry and exploit a large field of view to simultaneously
monitor many targets. 

The K2 mission plan is to point near the ecliptic,
sequentially observing fields as it orbits the Sun. This
observing strategy regularly brings new, well-characterized
target regions into view, enabling observations
of scientifically important objects across a wide range
of galactic latitudes in both the northern and southern
skies. K2 will perform a series of long, ecliptic-pointed
campaigns that use the proven {\it{Kepler}} infrastructure to
conduct new research into planet formation processes,
young stars, stellar activity, stellar structure and
evolution, and extragalactic science.
Herein, we present the details of our
new method of spacecraft operation and some early results
that characterize the scientific abilities of K2.

\section{The K2 Mission}

The K2 mission is driven by the spacecraft's
ability to maintain pointing in all three axes with only
two reaction wheels (Putnam \& Wiemer, 2014).
Figure 1 shows a schematic of the spacecraft and
the X-, Y-, and Z-axes. Solar pressure represents the only
disturbing force, which is controlled by wheels about
the Y- and Z-axes and by thrusters about the X-axis.
Sustained, stable pointing requires that the X-axis
disturbing force (roll axis) be minimized for extended periods.
This is accomplished by pointing in the orbital plane,
where the apparent motion of the Sun (caused by
the spacecraft's orbital motion while inertially pointed)
follows the spacecraft line of symmetry in the X-Y
plane over the course of a campaign, providing a balanced pressure and minimizing
the X-axis disturbance.

By carefully selecting the initial roll angle and correcting
for drift every 12 hours, the spacecraft can remain
stable in roll during the observation period. Meanwhile,
the reaction wheels control pointing about the Y- and
Z-axes as in the {\it{Kepler}} mission, absorbing the torque
generated by the solar radiation pressure. The accumulated
momentum is dumped through thruster
firings every two days. This operating mode provides
a fuel budget that allows for a 2-3 year mission duration.
Starting in October 2013, the operations team has
demonstrated and refined the techniques needed to
fly the spacecraft using this approach, perform an
on-orbit demonstration of performance, and deliver early commissioning science (\S5).

K2 has begun to observe a series of independent target
fields (campaigns) in the orbital plane (essentially the ecliptic
plane). The duration of each observing campaign
is limited by solar illumination constraints on the
spacecraft, bounded by power constraints on one
end and aperture illumination on the other. Science
observations are limited to about 75 days per field,
as illustrated in Figure 2. 

A generic K2 observing campaign timeline is shown in Figure 3.
Each K2 campaign will start when the target field comes
into view. A checkout period will be used to upload
target tables and configure the spacecraft attitude control
system for the upcoming campaign. The spacecraft will then be turned
to the new science attitude to collect pointing
alignment data, which are downlinked to the ground stations and to the project in near real-time.
Analysis of these data will then allow adjustments to the spacecraft configuration as needed
and start the science observations for the
campaign. To maximize the unbroken observation
period and reduce operational cost/complexity, all science data
will be stored onboard until the end of the campaign.
The spacecraft will then autonomously turn to point the high-gain antenna back toward the Earth
and the science data downlinked. Afterwards, the next campaign checkout period will began.
During science observations, in-situ health and safety
checks will be performed periodically and fault
protection will monitor spacecraft performance, placing
the spacecraft into safe-mode in the event of a
significant anomaly. 

K2 retains the {\it{Kepler}} point-and-stare observing
approach, minimizes operational changes, and utilizes
the data processing infrastructure with
modifications limited to those required to accommodate
the larger pointing drift and pointing control artifacts.
The quantity of the data delivered from each campaign will be far less than 
any {\it {Kepler}} quarter, as K2 will be observing only $\sim$10,000 targets, not the 150,000
observed in the nominal mission. Spacecraft communications will only slightly degrade over the anticipated 
K2 lifetime (2-3 years) and will have no effect on the science data quality discussed here.
All spacecraft operations are managed through
parameter table updates, and no flight software
changes are required. Aside from the failed reaction
wheels and the loss of two CCD modules (No. 3 in Jan. 2010 and No. 7 in Jan. 2013), 
the spacecraft has shown little performance
degradation and the remaining reaction wheels show
no signs of wear. Each K2 imaging science campaign will cover 
$\sim$105 square degrees and be self-contained. 

\section{K2 Key Science Goals}

The K2 mission was conceived to repurpose the
{\it{Kepler}} spacecraft within the limitations imposed by
two-wheel operations, and was influenced by the
large community response to a summer 2013 call
for white papers\footnote{
The community-produced white papers
are available at http://Keplerscience.arc.nasa.gov/TwoWheelWhitePapers.shtml.
}. K2 is a multi-field,
ecliptic-pointed mission that will allow observations
of thousands of targets covering the science areas
of transiting exoplanets, clusters of young
and pre-main sequence stars, asteroseismology,
AGN variability, and supernovae.

In each year of operation, K2 will observe
approximately 40,000 targets ($\sim$10,000 per field of view) spread over
four fields of view. K2 will collect data at
30-minute and 1-minute cadences and
will produce 80-ppm photometry for 12th
magnitude stars on 6-hour time scales (\S5).
K2 will observe in both the northern and
southern sky, in and out of the plane of
the Galaxy and, over two years, cover ten times more sky area
than the original {\it{Kepler}} mission.

The K2 mission will provide many opportunities for new discoveries
through its observation of targets and Galactic regions not
accessible to {\it{Kepler}} (e. g. Beichman et al., 2013). K2 is a community-driven
observatory with targets chosen from peer-reviewed proposals;
even the K2 fields will be chosen in coordination with the community (\S4).
As such, we can not know ahead of time the exact number or type of targets K2 will observe.
No predetermined set or type of targets is part of the mission concept.
The peer-review process will determine the actual target mix for each K2 field of view 
and thus what science will be possible from the observations made in each campaign.
Ecliptic fields which lie out of the Galactic plane are likely to be dominated by extragalactic targets while
those in dense star forming regions will probably receive proposals aimed mainly at young stars and clusters.
Additional science results are also expected
through observations of unique sources such as solar system objects, black-hole and x-ray binaries,
young massive stars, and numerous variable and pulsating stars.
K2 will use its unique assets to make observations 
capable of answering important questions in a number of science areas 
that are briefly highlighted below. 

\subsection{K2 Observations of Transiting Exoplanets}

{\it{Kepler}} discovered that
planets are common with small ones being plentiful (Howard et al., 2012).
To move from discovery to characterization,
host stars and their planets are needed which enable followup
yielding detailed properties. 
The K2 photometric precision will be lower and
the time per field shorter than the {\it{Kepler}} mission,
however, K2 has the potential to become a powerful exoplanet finder,
easily exceeding the capabilities of ground-based
surveys by large margins in sensitivity, field-of-view,
and continuous time coverage. Figure 4, based on early science results (\S5),
illustrates the ability of K2 to detect planets as functions
of stellar type and exoplanet radius.

\subsubsection{Observations of Exoplanets
Orbiting Low-Mass Stars}

M dwarfs offer a unique opportunity to progress from
planetary discovery to characterization. The proximity
of these host stars and the large photometric transit depths
allow a wide variety of additional observations
aimed at characterizing the atmospheres and properties
of such planets. For example, the exoplanet
GJ1214b (Charbonneau et al., 2009), while not in
the Habitable Zone or rocky, was discovered orbiting a
star relatively close to the Sun and lead to many 
interesting and useful follow-up studies. 

The highlighted box in Figure 4 shows the K2
discovery space for small rocky planets ($<$1.6\re) 
orbiting M dwarfs. Note that the rapid decrease
in stellar radius beyond spectral type M1V (\teff$<$
3600K) allows K2 to detect small planets even for the
generally faint population of M stars.

The results from the {\it{Kepler}} mission point toward a high
occurrence rate for small planets closely orbiting M
dwarfs (e.g., Dressing \& Charbonneau, 2013; Gaidos, 2013; Kopparapu, 2013). Due to their
low absolute luminosity, M dwarfs are a local population
and their distribution remains approximately
uniform across all K2 fields-of-view (Ridgway, et al.,
2014). K2 can observe approximately 4000
M dwarfs brighter than 16th magnitude per field. From
the {\it {Kepler}} estimated M dwarf small-planet frequency,
it is expected that K2 will discover
approximately 100 Earths and Super-earths (0.8$\le$R$_{Planet}$$\le$2.0) per
year($\sim$4.5 fields), with a few in the Habitable Zone (defined here as the empirical Habitable Zone; Dressing and 
Charbonneau, 2013). The larger
transit depths of M dwarf planets open the door to a
variety of ground- and space-based follow-up observations.
The detection of
such transits would permit detailed follow-up studies
as were done for GJ1214b, providing pre-launch targets for
JWST and the next generation of large-aperture
ground-based telescopes.

\subsubsection{Observations of Exoplanets Orbiting Bright Stars}

The main objective of the {\it{Kepler}} mission was to
measure the occurrence rate of planets around
Sun-like stars, particularly for Earth-size planets. {\it{Kepler}}
target stars were generally faint (V = 13 to 15) in order
to build a large enough sample that could be searched
continuously for four years. The {\it {Kepler}} results, that
exoplanets are common, is crucial for the design of
future instrumentation and missions that will make the
next leap in exoplanetary science - exoplanet characterization.
To accomplish this, the nearest and brightest stars,
harboring the most readily studied planets and
planetary systems need to be discovered. 

Bright stars (V$<$12) offer far more information than
simply telling us about the mere existence of a planet. 
A significant open question in exoplanetary science
concerns the interior structure and composition of
planets smaller than 2\re, irrespective of orbital location.
Most planets below 1.4\re~seem to be rocky
(Marcy et al. 2014), while planets larger than 2\re
~appear to be volatile rich mini-Neptunes. However,
the densities are known for only a handful of planets
between these bounds. 
 
Transits detected on bright stars by K2 will enable
precise Doppler spectroscopy, to provide planetary
masses and densities, and 
spectroscopic characterization of planetary atmospheric
properties. Bright stars are also amenable to 
high-resolution spectroscopy, high-resolution imaging, asteroseismology,
interferometry, proper motion, and parallax measurements, all useful 
to determine the host star properties

The number of bright dwarfs (V$<$12) that K2 can
observe in each campaign fluctuates throughout the
year from around 3000 to over 7000. Figure 5
shows the detectable planet sizes that can be found
in short-period orbits during a typical campaign. It
should be noted that about 40\% of the planet candidates
found by {\it{Kepler}} have periods less than eight days
and about half of these have sizes less than 2\re~ 
(Ciardi et al., 2013). Assuming 50\% of the targets are V$<$12 stars, K2 will observe approximately
20,000 bright stars in a given year and it should
find about 50 potentially rocky planet candidates 
based on {\it {Kepler}} statistics. The ecliptic is already home to 
a number of bright exoplanet host stars: 15 are bright (V=5-8) with short-period
RV planets and 23 are transiting systems.

\subsection{K2 Observations of Open Clusters}

MOST, CoRoT and {\it{Kepler}} have all made contributions
to open-cluster science, but for limited
samples and for distant, faint and crowded clusters.
K2 campaigns (\S4) are planned to survey a rich area of the sky, containing the
brightest and best characterized clusters (see Table
1). Much is already known about these clusters
and associations, including cluster membership
and stellar properties. 

In these clusters, K2 can survey a few thousand bright pre- and early main
sequence stars down to 0.15\msun. Assuming a
photometric precision of 80 ppm on transit timescales
for a 12th magnitude star, K2 can comfortably detect
planets down to 2.5\re, the size of Super-Earths, in
all the clusters listed in Table 1. The incidence
of short-period planets around field stars (Fressin et
al., 2013; Petigura et al., 2013) suggests that a few
transiting hot Jupiters and Neptunes will be found
per campaign, as well as several tens of Super-Earths
and smaller planets assuming $\sim$1500 cluster members are observed (see \S3.1). 
To date, only a handful of planets have been detected in
open clusters (Quinn et al., 2012; Meibom
et al., 2013; Brucalassi et al., 2014). These discoveries
indicate that the incidence of large planets in
clusters is similar to that around field stars. The K2
mission has the ability to determine if small
planets are as common in open clusters as in the field.
Planetary systems
discovered in the Hyades will be particularly valuable
for follow-up studies with JWST, being bright and only
46 pc from the Sun.

The past decade has seen an increase
in the number of rotation-period measurements
available for pre-main sequence and early main
sequence stars, but theoretical models still struggle
to reproduce all the available data (e.g., Gallet and Bouvier, 2013).
Typical rotation periods
for young cluster stars range from 1 to 20 days (Irwin \&
Bouvier, 2009), and have large and distinct modulation
amplitudes, easily detectable and distinguished
from transit events. K2 observations of many cluster members can
provide an essentially complete rotation census
in each targeted cluster. 

Detached, double-lined eclipsing binaries provide
model independent determination of the masses,
radii, effective temperatures, and luminosities of
both stars from the light and radial velocity curves
of the system. 
CoRoT observations of the open cluster NGC2264 
(Gillen et al., 2013) and {\it {Kepler}} observations of NGC 6811 (Janes et al., 2013)
can be used to estimate that K2
will discover $\sim$10 new eclipsing binaries
per cluster for which the masses and radii of both
components can be determined to about 1\%. The
continuous sampling achievable by K2 also 
enhances the sensitivity to moderate period
(i.e., 8-20 days) eclipsing binaries, enabling their
(spin) evolution to be studied as a function of the
mutual interaction between the two stars. 

The use of asteroseismology 
can significantly advance our understanding
of stellar evolution, stellar-interior physics,
and stellar populations (Chaplin \& Miglio 2013). Nowhere will this be more
powerful then when used to combine high-precision
photometric observations with strong prior information on
age, distance, and metallicity in clusters that span a wide range in
mass, chemical composition, and evolutionary state.
K2 observations of bright stars within clusters, 
e. g., O-B stars (Aerts et al., 2003,2013; Degroote et al., 2010) or A and F pulsators
(Michaud \& Richer, 2013) can yield results
on radiative levitation and element depletion processes, and the onset of
near-surface convection, and
allow comparisons of distance scales for various cluster stars
using the standard pulsating variables such as 
RR Lyrae stars or Cepheids. 

Studies of Galactic star populations with K2, both in clusters and in the field,
could measure radii, masses, distances, and ages of
thousands of giants providing a unique asteroseismic
survey spanning a wide range of vertical and
radial structure in our Galaxy. These data would allow
Galactic mass and age gradients to be studied and
characterized for a wide range of populations, out
to similar distances to those probed by red-giants in
the {\it{Kepler}} field. Detections in the {\it{Kepler}} and CoRoT
fields are already being utilized (e.g., see Miglio et al., 2013; Stello
et al., 2013), but their fixed pointings carry obvious limitations that
will be lifted for K2 (Figure 6).  Combined with astrometric data
from Gaia, K2 observations of field red giants would provide a valuable community dataset
(Chaplin \& Miglio, 2013).

\subsection{Observations of Star-Forming Regions}

Understanding how giant planets form and accumulate their gaseous envelopes is still an
observational frontier. K2 campaigns will study star-forming regions providing 
the opportunity to 
test planet-formation theories by probing planetary properties during formation or
immediately thereafter. Standard core accretion (e.g., Lissauer et al., 2009) predicts that
proto-giant planet sizes are large (50-100 \rjup) during the slow-gas accretion phase.
After accretion has ended, the planet contracts to a more modest size ($\sim$1.1 to 1.6\rjup) 
for a 1 \mjup~planet. The actual value of the planetary radius at this young age is
extremely sensitive to how efficiently the planet radiates away accretion energy. Measuring small
planet sizes with K2 will show that the radiation process is extremely efficient and this energy is
easily lost: planets start off their life with a "cold start". Detecting much larger planets with K2
might indicate an earlier phase of evolution or that the accretion process is not efficient in
radiating energy away.

The migration of hot Jupiters early in a star's life (less than 1 Gyr) is believed to result from
one of two mechanisms: migration within the protoplanetary disk ("disk migration"; e.g., Raymond
et al., 2006) or some mechanism that dynamically produces highly eccentric orbits, evolved via
tides ("dynamical migration"; e.g., Albrecht et al., 2012). These two scenarios deliver the hot
Jupiters at different times during the star's evolution, promptly, or delayed to greater than 0.5
Gyr by tidal evolution. Therefore, any K2 measured orbital periods of Jupiter-mass planets will be a
sensitive test of hot vs. cold-start models 
and migration scenarios (see Marley et al. 2007, Spiegel and Burrows, 2012).
Characterizing planetary
systems at the few Myr age of T Tauri stars will explore
the dynamical evolution of planetary systems
by comparison with the results of more
mature planetary systems. Measured abundances
of hot Jupiters around young stars can be compared
with measurements obtained by {\it{Kepler}} for older
field stars.

The variability of the young stars
is likely to be higher than field stars (Van Eyken et al.,
2012), which will provide interesting science in itself,
but may impact transit detection and RV follow-up. 
However, 
hydrodynamics studies of forming giant planets
(e.g., Klahr \& Kley, 2006; Gressel et al., 2013)
suggest that these objects may be more readily
detectable than previously believed. For example,
in their 3-D radiation hydrodynamics study, Klahr
(2008) found that the nascent planet is surrounded
by a warm pressure-supported bubble with a
large pressure scale height (H/R$\sim$0.5) rather than
a thin Keplerian circumplanetary accretion disk (H/R
$\le$ 0.1). If such gas/dust bubbles are indeed signposts
of giant planet formation, their large scale
height would likely make them detectable through
a transit signature as they occult the central star.
K2 observations of a very large sample of
well-characterized young stars during their early formation
processes would yield valuable information as well as targets for follow-up with ALMA.

\subsection{K2 Observations of Variable Extragalactic Sources}

{\it{Kepler}} observations of extragalactic sources have
allowed the initial exploration of AGN variability
on a variety of time scales (hours to months)
(Mushotzky et al., 2011). Edelson et al., (2013) and
Olling et al., (2013) have found new AGN exhibiting
low levels of activity, and have discovered at
least four supernovae. K2 can provide observations
of thousands of galaxies, many well studied (e.g.,
COSMOS field), through its observations at high
Galactic latitude, areas rich in extragalactic sources.

\subsubsection{AGN Variability}

Theoretical models (Arevalo et al., 2009; Breedt et
al., 2010) suggest that the power spectral densities
(PSDs) of accretions disks have spectral indices of
-1.4 to -2.0. However, {\it{Kepler}} observations provide
a few well-determined PSDs yielding slopes of
-3.0 (Mushotzky et al., 2011). The steep measured
slopes are inconsistent with model predictions, but
are based on a small sample. In addition, BL Lac
microvariability (0.5-1\%) measurements, such as
for W2R1926+42 observed with {\it{Kepler}}, provide
evidence for periods of strong flaring and periods of
relative quiescence. This type of behavior cannot be
fit with the existing simple models for Seyfert galaxies.
A larger sample of sources is needed, including
various AGN types with varying properties such as
luminosity and black-hole mass.

Edelson and Malkan (2012) list 4316 AGN in the
entire sky brighter than J=16 (R$\sim$17) suggesting
nearly a dozen bright AGN will reside in each K2
field. It is likely that K2 will observe over a hundred
microvariable extragalactic sources during its mission (Edelson et al.,
2013), allowing a broad range of variable AGN to
be observed in statistically significant numbers that will
provide robust testing of the current models.

\subsubsection{Progenitors of Type Ia Supernovae}

Over 20,000 known galaxies brighter than 19th magnitude
will be visible per year in K2 fields. K2 observations could allow
detection of supernovae for the observed galaxies. Taking
the typical observed supernova rate (0.7 supernova/century/
galaxy; Olling et al., 2013), K2 will observe many galaxies for a variety of scientific reasons, some of them,
a few tens per year, will contain supernova whose early light curves will be serendipitously collected.

Olling et al. (2013) present a Type Ia
supernova light curve obtained with {\it{Kepler}}. The continuous
light curve coverage starting
prior to onset yields unique observations of the early
event. This early rise-time information can, for all
types of supernovae, tightly constrain the progenitor
of the explosion and detail the subsequent shock
physics (Woosley et al., 2007; Kasen, 2010; Haydon
et al., 2010). Theoretical models (Nakar \& Sari, 2010;
Kasen, 2010) show that a single degenerate star will
cause smooth shock emission in the first few hours
to days after the explosion, while double degenerates
are expected to brighten monotonically. There are no
existing or planned facilities that can provide initial and
early rise-time information for supernovae light curves
with the precision and time coverage available with K2.

\subsection{K2 Micro-Lensing Observations}

For the majority of the planets currently detected
by microlensing, a parallax determination 
is needed to determine their
masses and distances (Gould, 1999; Dong, 2009).
K2 might provide this missing ingredient by measuring
"microlens parallaxes" (Gould \& Horne, 2013).
Planets detected in this manner are particularly interesting because
they inhabit a cold region that is inaccessible to
other detection methods (Gould \& Loeb, 1992).

The K2 solar orbit and wide field-of-view make it
well suited to measure microlens parallaxes. K2
will see a radically different event from the Earth-based
event because it is displaced by approximately
0.5 AU. When combined with
ground-based observations, the measured parallax yields
the host star and planet masses. K2 could observe as many as
12 microlensed
planets in a single campaign (Gould \& Horne, 2013),
enabling the first robust measurement of the mass
function of cold, low-mass planets across different stellar populations. 
One of the most spectacular results from microlensing
is the detection of "free-floating" planets. Here too,
the microlensing parallax will measure the mass of
these objects and confirm their planetary nature.

K2 microlensing results would also provide important input to future
microlensing surveys, in particular WFIRST, by
estimating the expected yield determining the resources needed
for WFIRST follow-up. 
Field 9 (\S4, Figure 7), is a forward looking field, and is ideal for a microlensing study as it is
near the Galactic center. Forward looking fields, that is pointing in the positive velocity vector of the spacecraft,
offer the ability of simultaneous Earth-based observations during the K2 observations. However, the Earth and moon enter into 
the K2 field of view during the 75-day campaign. Our initial planning for Field 9 is likely to consist of shorter visits
to dense star fields and to use full-frame readouts of the entire array instead of individual target pixels.
This unique K2 campaign will be dedicated to a microlensing investigation and will operate differently than the other campaigns.
In order to obtain the highest scientific return from this microlens study, the exact method of field 9 observations will be 
planned in coordination with the microlensing community.

\section{K2 Community Involvement: Field and Target Selection \& Science Archives}

Based upon the the spacecraft abilities, the two-wheel white paper call 
and additional community feedback, a series of ecliptic K2 fields of view have been 
proposed (Figure 7). These fields are rich
in the types of targets requested by the community. Table 2 lists the planned K2 campaigns, the RA and DEC of the center of the
field of view, and the date of mid-campaign. The locations of 
Fields 1 and 2 are now fixed with Fields 3-9 still accepting community input.
Each K2 field position needs to be finalized approximately six months prior to its start.
The location of each campaign field attempts to optimize the science yield
while balancing the spacecraft abilities as well as providing the requisite number of guide stars for the telescope.
Maximizing the science, for example, might move a field start or end date a few days to include more stars from a highly
proposed cluster. These types of trades in the field position are balanced with the 
mission concept of providing long-term observations (75 days) at each field and moving from field to 
field with as few as possible non-science collection days in between. Details of the location of each field 
and the method for user input can be found at the Kepler science Center$^{13}$. 

K2 will observe 10,000-20,000 targets per
campaign collected on a 30-min cadence and an
additional 100 targets per campaign collected on
a 1-min cadence. Target numbers are limited by
on-board storage and compression, downlink capability,
and campaign length. The precise number of
targets observed will vary by campaign and will be
an optimized trade based upon targets and spacecraft
pointing performance. The scientific community
will propose all targets for each campaign. 
The K2 mission will provide a catalog of potential sources within each campaign field based on
photometric and astrometric surveys such as Hipparcos, Tycho-2, UCAC-4, 2MASS and SDSS. The
catalog will contain celestial coordinates, proper motions, parallaxes, broadband magnitudes,
inferred K2 bandpass magnitudes, as well as estimates of stellar properties (e.g., effective
temperature, surface gravity, and metallicity).
Each catalog will be archived at the MAST prior to calls to the community for target proposals.
The K2 targets discussed herein and the potential results from our listed key
science areas are subject to successful proposals from the community.
There will be no restrictions on what type of science may be proposed 
and no exclusive use period for any K2 data. 

K2 has begun making preliminary science observations
starting on 8 March 2014 (Campaign 0), that was moved slightly to include the open cluster M35.
Of the $>$100,000 targets proposed for Campaign 0, approximately one-third (with many duplicates) 
proposed targets in M35 with the remaining majority being bright stars, galaxies, and M stars. 
All K2 campaigns will have peer-reviewed target selection; Campaign 0 and 1 targets have already been solicited and 
submitted with campaign 1 targets under review at present 
via a project selected external review panel. Future campaign targets would be solicited through 
a NASA announcement of opportunity (AO) similar to other NASA guest observer (GO) programs with the proposals
reviewed by a NASA appointed
external panel. The details of this review process will be formulated 
and a guest observer (GO) program is anticipated to begin in summer 2014 if the K2 mission is approved.

The K2 mission will deliver all time-series data to
a legacy archive hosted at the Mikulski Archive
for Space Telescope (MAST; http://archive.stsci.
edu/kepler). The K2 archive will 
reuse the architecture of the {\it{Kepler}} archive,
with the advantage that it will look and
feel much like the {\it{Kepler}} archive. Data products
will be produced and archived three months after
the end of each campaign, allowing the community
to rapidly pursue ground-based follow-up and
compete for NASA and other funding resources as early as possible.
For Campaigns 0-2, calibrated pixels (bias corrected, flat fielded, smear corrected, and with ADU's converted into electrons) 
will be delivered on this nominal schedule, while the higher-level products will be delivered at a later date in
2014. In support of community-led exoplanet candidate detection, vetting,
validation, verification and follow-up, the Exoplanet
Archive (http://exoplanetarchive.ipac.caltech.edu)
will host the list of potential exoplanet transit events
and associated diagnostic data, as
well as provide tools and additional data resources
for their exploitation.
K2 community support will be provided by the 
{\it{Kepler}} Science Center at the NASA Ames
Research Center\footnote{http://Keplerscience.arc.nasa.gov/K2}.

\section{K2 Early Science Results}

Based upon a series of spacecraft tests
between October 2013 and February 2014,
spacecraft operations were refined to
maximize two-wheel performance from K2 and deliver science verification results. 
These tests have demonstrated all of the functionality required
for the K2 mission. Science validation data were
collected to characterize K2 mission performance, and are presented here
to provide a demonstration of K2's ability to carry out its key science goals.

The K2 photometric precision primarily depends on motion
of the spacecraft boresight during timescales shorter
than an exposure and timescales longer than a single exposure due to motion caused by
solar-induced drift. 
Tests to date have demonstrated spacecraft
jitter (high frequency motion) performance comparable to {\it{Kepler}} over 30
minutes, or one exposure. The measured Full-
Width Half Maximum of the K2 point spread function,
a measure of spacecraft jitter, is within 5\% of
the fine-point {\it{Kepler}} point-spread function across
the entire field-of-view. Figure 8 provides a fit of
the {\it{Kepler}} point-spread function to a K2 target close
to the spacecraft boresight. Spacecraft jitter during
two-wheel operation is therefore generally only a
few percent larger than in the three-wheel {\it{Kepler}}
mission and is not a major concern for K2 photometric
precision.

The remaining component of photometric precision
is solar-induced drift. This low-frequency motion due
to solar pressure and subsequent thruster firings
causes targets to drift across detector pixels and is
the dominant factor in photometric precision after
photon statistics. Thrusters are used to manage
momentum by periodically resaturating the reaction
wheels and to correct for solar-induced roll-angle
drift. 
Thruster firings kept targets localized to within three
pixels during early tests, and later testing demonstrated
1-pixel pointing using focal plane mounted fine guidance sensors.

For the purposes of providing a photometric precision
measure that can be compared with the {\it{Kepler}}
mission, an analysis was conducted on a typical
uncrowded 12th magnitude K2 target. 
The parameters listed here for {\it {Kepler}}
are not identical to the baseline mission photometric values specific to 
transit searches as presented in Koch et al. (2010).
Here, we make use of the publicly available
community software tools. 
To make our comparison, an identical
analysis was performed on a {\it{Kepler}} target of the
same magnitude. Both light curves are shown in
Figure 9. The standard deviations of these two
light curves, after normalization by a low-frequency
filter, are listed in Table 3.

To compare K2's transit detection sensitivity with
{\it{Kepler}}'s, motion systematics and stellar variability
are removed from these time series using a standard
48-hour Savitzky-Golay filter (Savitzky \& Golay, 1964).
The results are shown in Table 3 for photometric precision over
a 6-hour exoplanet transit duration ($\zeta$). {\it{Kepler}} operational
experience indicates that 7$\zeta$ was a reasonable
threshold for transit detection and is used for
all K2 planet yields estimated above.

Figure 10 provides a comparison of K2's and
{\it{Kepler}}'s photometric precision as a function of target
magnitude. The metric quantifies sensitivity to 6-hour
transits and was calculated consistently for both
target series. The comparison indicates that current
K2 performance is within a factor 3-4 of {\it{Kepler}}'s
precision. The {\it{Kepler}} sample consists of quiet G
dwarfs, selected randomly across the field of view.
The K2 sample is also selected randomly across the
full field-of-view, however the nature of each target is
unknown; it is unlikely they are all quiet dwarfs, resulting
in the scatter observed in the distribution.

As part of an early science demonstration with K2, WASP-28 was 
observed for 2.6 days in short cadence (1-min sampling) mode during an engineering 
test in January 2014.
WASP-28 is a 12th magnitude, Sun-like star with a somewhat subsolar
metallicity and known to host a
Jupiter-sized exoplanet with an orbital period of 3.4 days (Anderson  et al. 2014).

The K2 light curve (Fig. 11) was extracted from calibrated pixel data using an 
elliptical aperture that was allowed to recenter its position between exposures. 
In each exposure, we removed a local background determined from the median flux 
in nearby pixels. Finally, we passed the data through a median filter with a 1-day window.

Near the end of our short observation, we observed a single transit 
of the planet WASP-28b as it moved across the host
star. We fit a limb-darkening transit model (Mandel \& Agol 2002) to our observations
and obtained planet parameters
consistent with published values (Anderson  et al. 2014).
The residuals of the transit fit to the data show a point-to-point (rms) scatter of 0.16\%. This
equates to a 6-hr integrated noise level of 84 ppm. 

\section{Conclusion}

K2 is a new mission both in spacecraft control and science opportunity.
It makes use of the existing {\it Kepler} spacecraft and the large-area focal plane array
to provide high-precision, long-duration photometric observations allowing
new scientific discoveries in many areas of astrophysics. K2 is currently undergoing Campaign 0
and, if approved by NASA, is planned to continue observing 
a new field approximately every 3 months starting around 1 June 2014.
K2's photometric precision and observing ability will far 
exceed ground-based telescopes and enable
discoveries of high-value transiting exoplanets, a variety of aspects in 
stellar evolution, and new windows to explore in extragalactic science.
Observing 4 to 5, 75-day campaigns each year, 
K2 will scrutinize the sky along the ecliptic plane.

K2's key science objectives were laid out above (\S3) and we have presented 
early science results (\S5) attesting to the mission's ability to carry out these goals.
K2 is a low-cost space astrophysics mission capable of elucidating many aspects in stellar astrophysics across
the H-R diagram and providing detailed observations of variable galaxies and early time
observations of supernovae. In addition, K2's unique exoplanet survey will fill the gaps in duration and sensitivity between 
the {\it{Kepler}} and
TESS missions, and offers pre-launch exoplanet target identification for JWST transit spectroscopy.

\acknowledgements{We thank Andy Adamson, Michael Bicay, Padi Boyd, Alex Brown, Jessie Dotson, Nick Gautier, 
Doug Gies, Richard Green, Steve Kawaler, Caty Pilachowski, and Dave Silva for reading early drafts and
making comments that led to a better paper.
C. Aerts, S. Basu, T. Bedding, K. Brogaard, J. Christensen-Dalsgaard, S. Kawaler, H. Kjeldsen, D. Kurtz,
and D. Stello are acknowledged for making valuable contributions to the work involved in this paper.
The authors also wish to thank Wendy Stenzel for producing the graphics associated with this
paper and Mark Messersmith for keeping us in line during the work.
Ball Aerospace is gratefully acknowledged for their long-term commitment to {\it Kepler} and now K2.
}

{\it Facilities:} {\it{Kepler}}, K2

\begin{deluxetable}{lcccl}
\tablenum{1}
\tablecolumns{5}
\tablewidth{0in}
\tablecaption{K2 Open Clusters}
\tablehead{
\colhead{Cluster} & \colhead{Age (Myr) } & \colhead{Distance (pc)} & \colhead{K2 Campaign} & \colhead{Ref.} \\
\colhead{} & \colhead{} & \colhead{} & \colhead{(Proposed)} & \colhead{}
 }
\startdata
Taurus & 2 & 140 & 4 & Rebull et al., 2010 \\
Upper Sco & 10 & 130 & 2 & Pecaut et al., 2012 \\
M21 & 12 & 1200 & 9 & Piskunon et al., 2011 \\
M18 & 32 & 1300 & 9 & Santos-Silva \& Gregorio-Hetem, 2012 \\
M25 & 92 & 620 & 9 & Piskunon et al., 2011 \\
M35 & 100 & 800 & 0 & McNamara et al., 2011 \\
M45 & 125 & 135 & 4 & Bell et al., 2012 \\
NGC 1647 & 150 & 547 & 10 & Piskunon et al., 2011 \\
NGC 6716 & 150 & 547 & 7 & Piskunon et al., 2011\\
Hyades & 630 & 46 & 4 & Schilbach \& Roser, 2012 \\
M44 & 630 & 160 & 5 & Boudreault et al., 2012 \\
M67 & 4300 & 908 & 5 & Dias et al., 2012, \\
\enddata
\end{deluxetable}

\begin{deluxetable}{cccc}
\tablenum{2}
\tablecolumns{4}
\tablewidth{0in}
\tablecaption{K2 Campaign Fields} 
\tablehead{
\colhead{Field} & \colhead{RA} & \colhead{DEC} & \colhead{Date of Mid-Campaign} 
 }
\startdata
0 & 06:47 & +21:23 & 4 May 2014 \\
1 & 11:38 & -01:11 & 22 Jul 2014 \\
2 & 16:34 & -22:49 & 14 Oct 2014 \\
3 & 22:21 & -11:37 & 5 Jan 2015 \\
4 & 03:46 & +18:08 & 29 Mar 2015 \\
5 & 09:19 & +14:12 & 20 Jun 2015 \\
6 & 14:01 & -13:16 & 11 Sep 2015 \\
7 & 19:34 & -22:38 & 3 Dec 2015 \\
8 & 01:04 & +05:12 & 24 Feb 2016 \\
9 & 18:24 & -24:12 & 17 May 2016 \\
\enddata
\end{deluxetable}

\begin{deluxetable}{lcc}
\tablenum{3}
\tablecolumns{3}
\tablewidth{0in}
\tablecaption{Comparison of {\it{Kepler}} and K2 Photometric Performance}
\tablehead{
\colhead{Parameter\tablenotemark{a}} & \colhead{{\it{Kepler}}} & \colhead{K2} 
 }
\startdata
Standard Deviation (1$\sigma$) & 99 ppm & 404 ppm \\
6-hour phot. precision ($\zeta$) &  18 ppm & 82 ppm \\
\enddata
\tablenotetext{a}{Parameters determined as described in \S5. {\it {Kepler}} baseline photometric values are described in Koch et
al. (2010).}
\end{deluxetable}

\clearpage

\begin{figure}
\includegraphics[angle=0,scale=1.0,keepaspectratio=true]{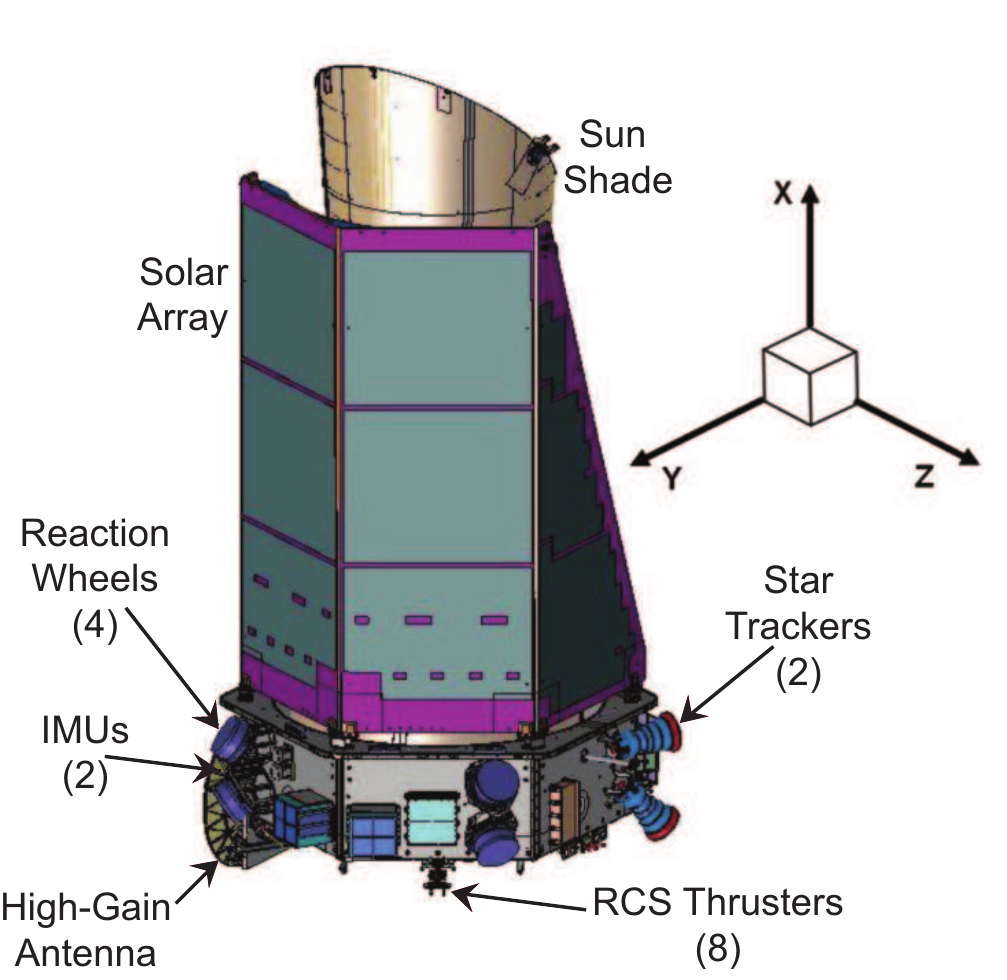}
\caption{
{\it{Kepler}} spacecraft coordinate system.
}
\end{figure}

\begin{figure}
\includegraphics[angle=0,scale=0.8,keepaspectratio=true]{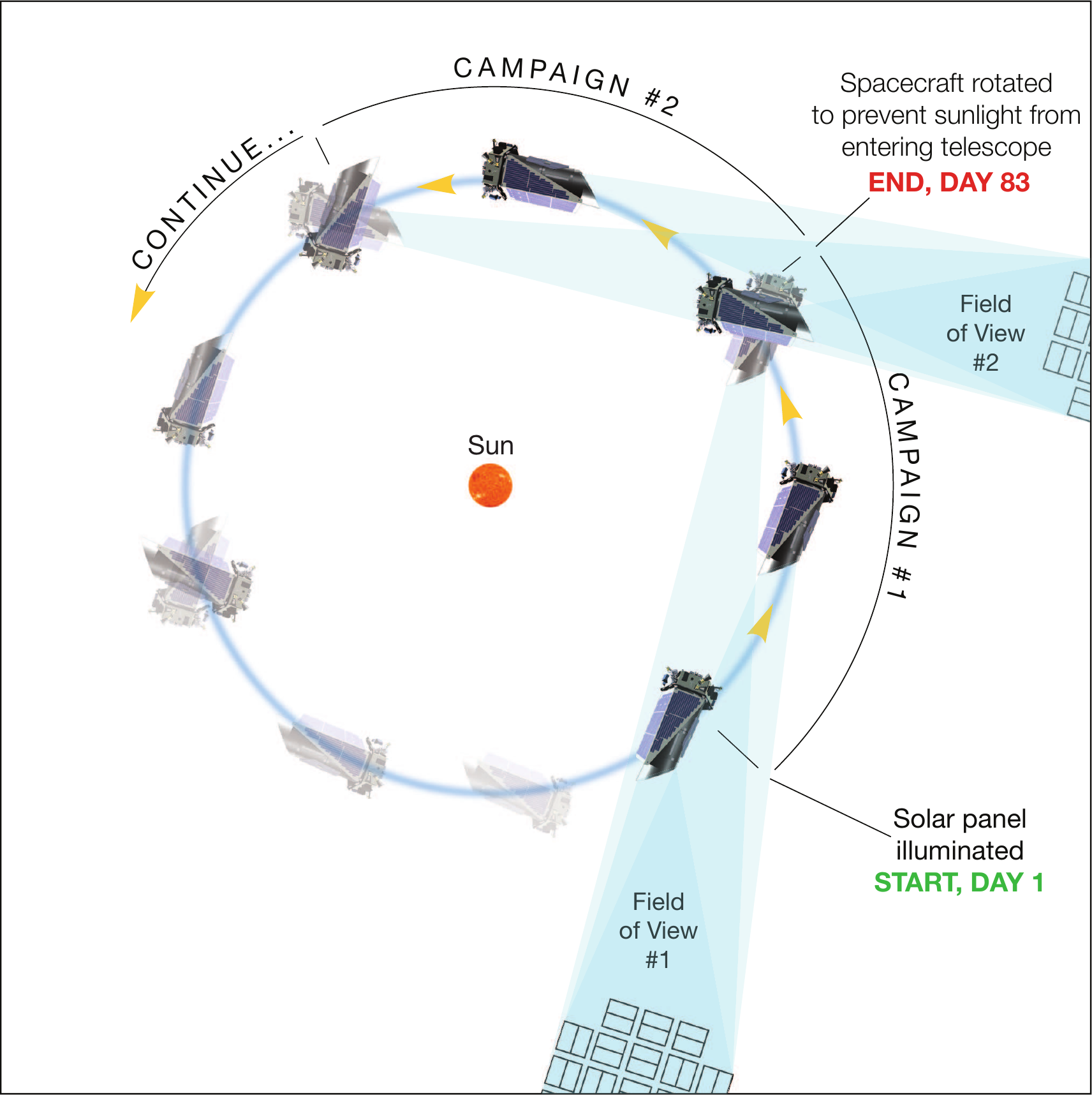}
\caption{
The K2 mission will observe sequential ecliptic
campaigns with a duration of $\sim$83 days, where 75
days are dedicated to science.
}
\end{figure}

\begin{figure}
\includegraphics[angle=0,scale=1.0,keepaspectratio=true]{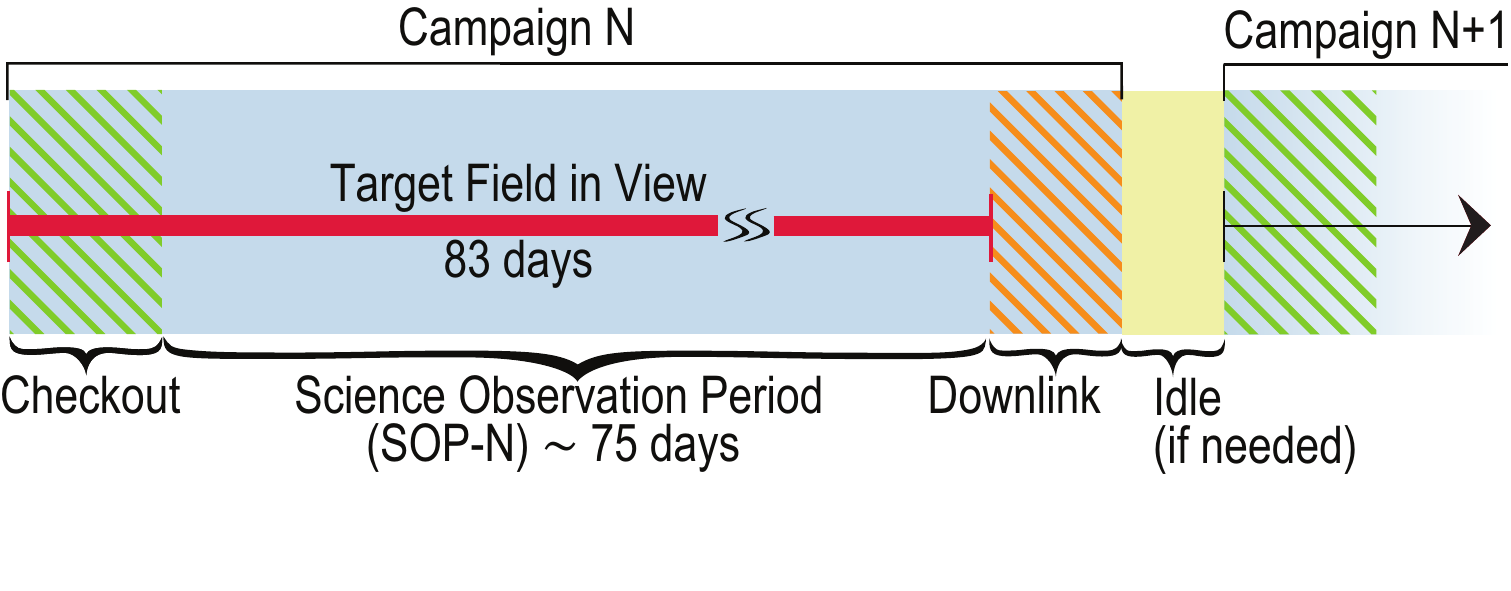}
\caption{
Schematic representation of a typical K2 campaign showing how each will provide a long, uninterrupted
science observation period. An initial checkout
period assures precise initial pointing, while all data are
returned at the end of the campaign during an approximately 2-day downlink period. Idle time can be
inserted between campaigns to allow flexibility in choosing
the next field-of-view.
}
\end{figure}

\begin{figure}
\includegraphics[angle=0,scale=1.0,keepaspectratio=true]{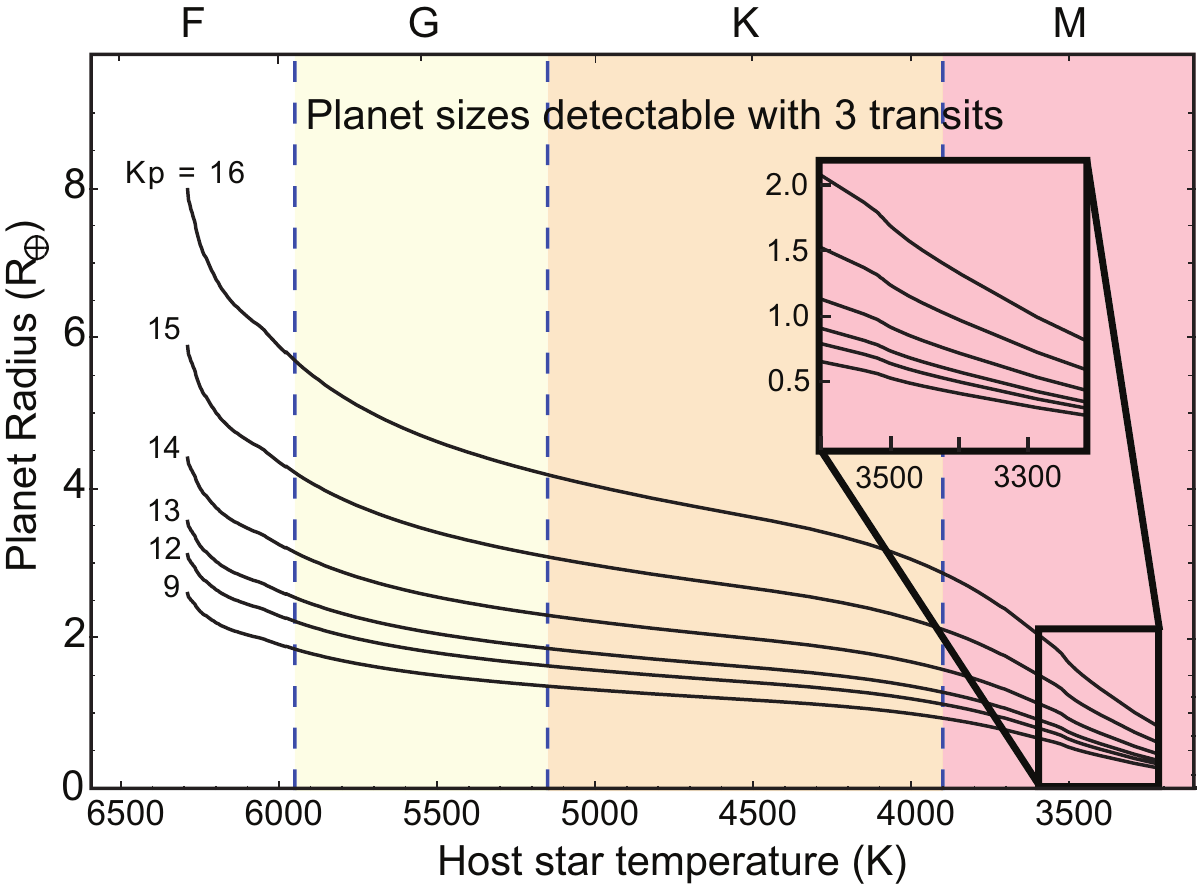}
\caption{
K2 will detect small, short-period
planets around cool dwarfs. Using three transits
as a metric for detecting short-period candidate
exoplanets (based on K2 photometric performance
as described in \S5) the planet-detection thresholds
for various dwarf-star types are shown for a range of
{\it{Kepler}} magnitudes (Kp $\sim$ R-band). For high 
signal-to- noise events, two transits are sufficient to support
the detection of small Habitable Zone planets orbiting
M dwarfs.
}
\end{figure}

\begin{figure}
\includegraphics[angle=0,scale=1.0,keepaspectratio=true]{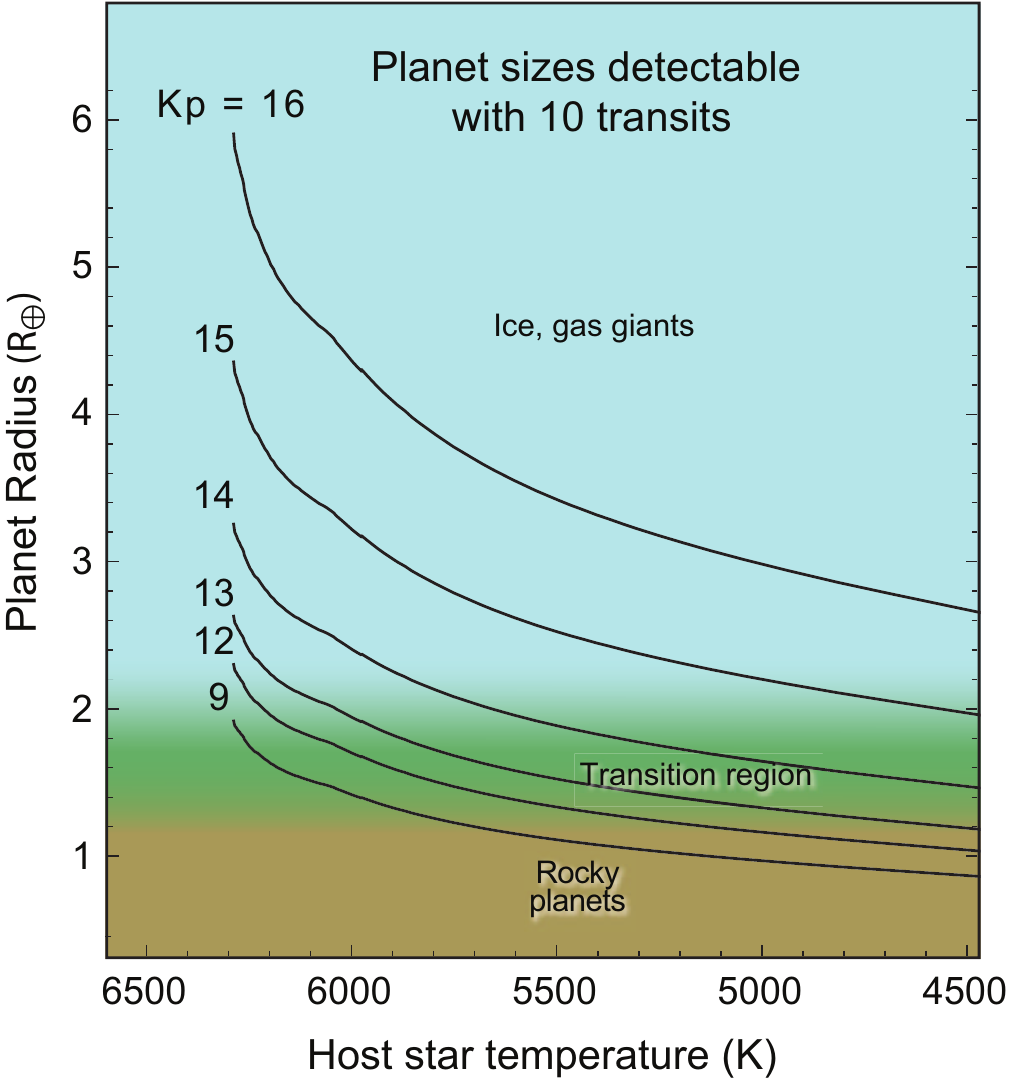}
\caption{
K2 will discover exoplanets which probe the transition and
rocky regimes for G dwarfs (to 13th magnitude)
and for cooler stars. Based on {\it{Kepler}} statistics
and the expected photometric performance(\S5),
K2 is expected to detect $\sim$50 planet candidates per
year with orbital periods $<$8 days and radii $<$2\re. 
}
\end{figure}

\begin{figure}
\includegraphics[angle=0,scale=1.0,keepaspectratio=true]{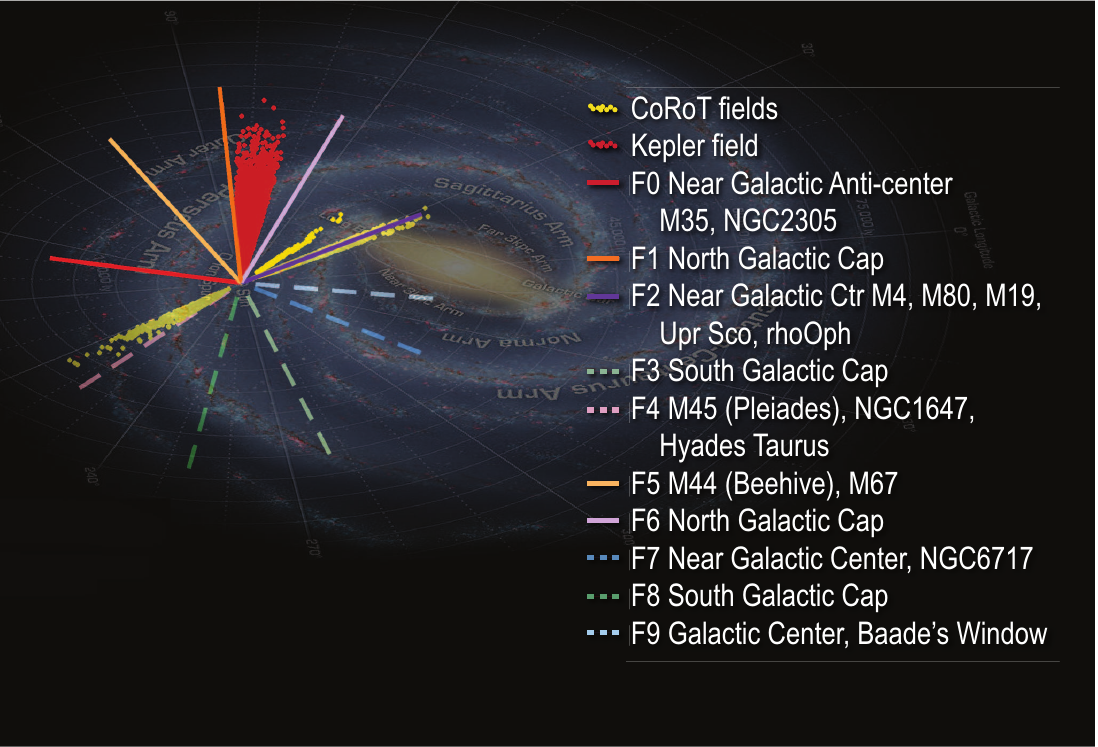}
\caption{
The Galactic distribution of oscillating red giants accessible
to K2 is compared to those observed by {\it{Kepler}} and CoRoT.
In this cartoon view of the Galaxy, we see that K2 fields will increase the distribution 
of environments in which asteroseismology can be applied in order to measure the masses, radii,
distances, and ages of stars. 
}
\end{figure}

\begin{figure}
\includegraphics[angle=0,scale=0.7,keepaspectratio=true]{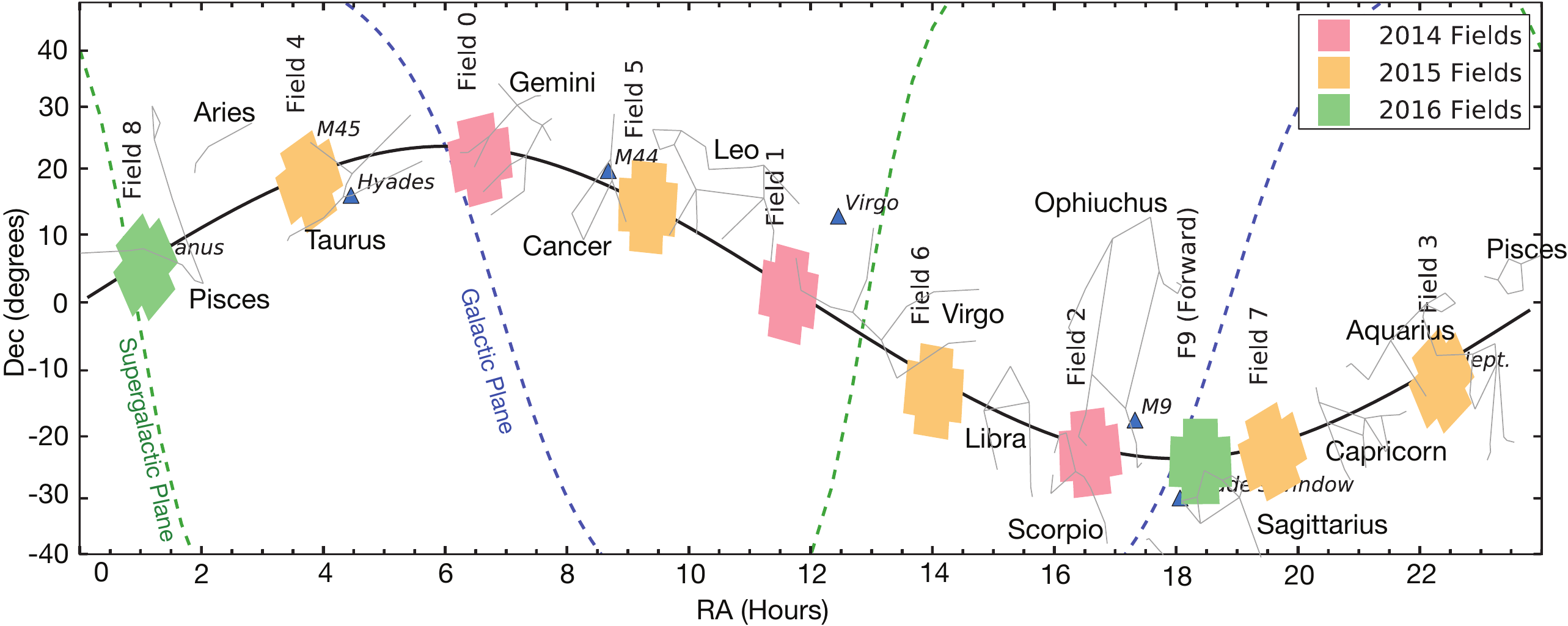}
\caption{
The K2 ecliptic observing field sequence was anchored by Field 4, which collects observations of the
Pleiades and Hyades clusters in 2015 (see Table 2). Earlier and later fields march around the ecliptic in steps of
83 days as described in Figure 2.
Field 9 is designed to observe along the spacecraft velocity vector (pointing toward the Earth),
instead of standard the anti-velocity vector, to provide a unique K2 survey of Baade's window 
to be used for a microlensing study. This Earth-pointing orientation was chosen to allow
simultaneous ground-based support.
}
\end{figure}

\begin{figure}
\includegraphics[angle=90,scale=1.0,keepaspectratio=true]{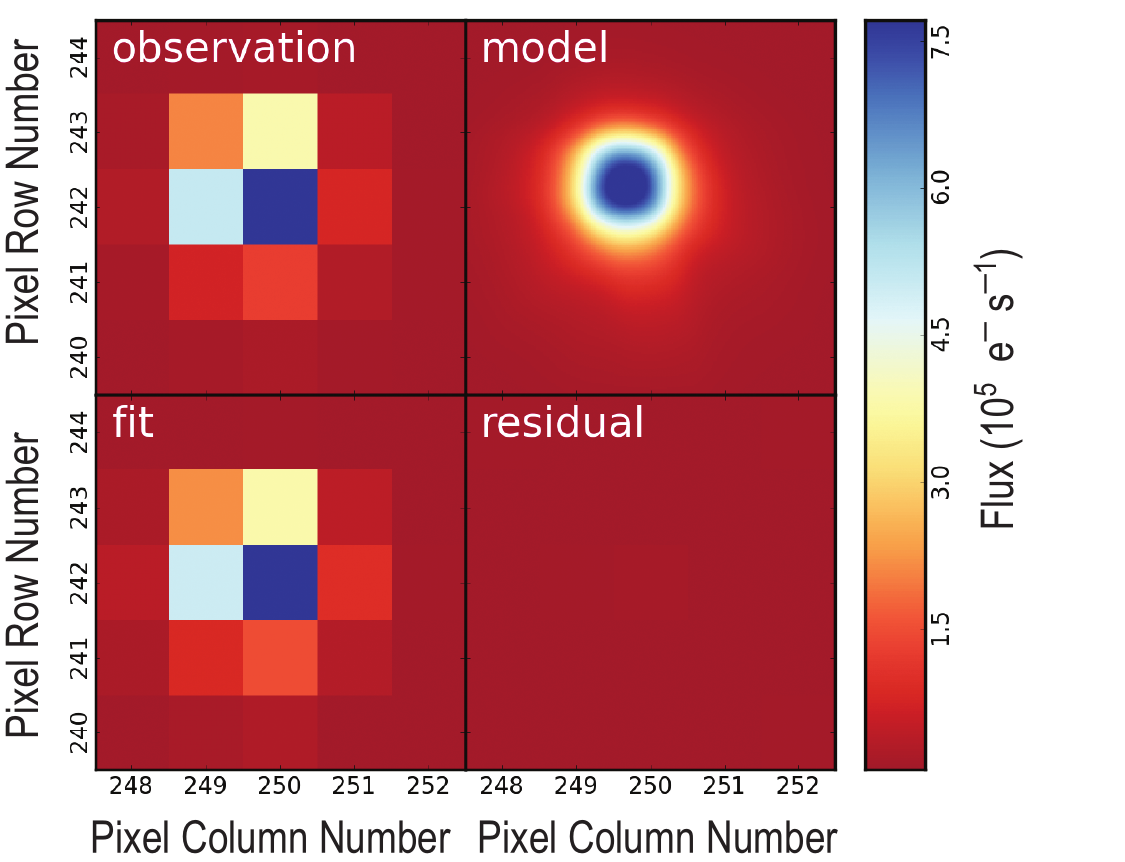}
\caption{
The K2 Point-Spread Function
(PSF) is well fit by the empirical {\it{Kepler}} PSF
model. Increased high-frequency movements (jitter) during
two-wheel operations increases the {\it{Kepler}} PSF
FWHM by 2-5\% across the detector. From
top-left to bottom-right: a 30-minute K2 observation
of a 12th magnitude star, the best-fitting
{\it{Kepler}} PSF model, the best-fit model binned
over detector pixels, and the fit residual.
}
\end{figure}

\begin{figure}
\includegraphics[angle=0,scale=0.6,keepaspectratio=true]{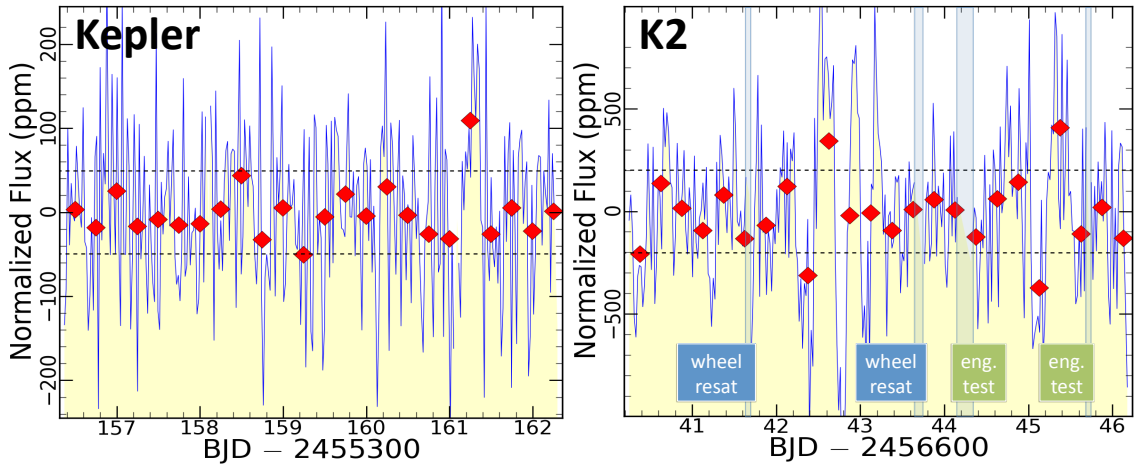}
\caption{
The December 2013 on-orbit K2 test demonstrated a photometric precision of 82 ppm for a 12th
magnitude point source over 6-hour exoplanet transit timescales. This estimate was based on the K2
light
curve in the right-hand panel. The dashed line represents the 1$\sigma$ point-to-point 
(30-min) standard deviation
of all data points. The red points are 6-hr averages of the data. The shaded regions represent
data-collection
periods where reaction wheel resaturations and fine-guidance tests made the data unusable. For
comparison, a
similar plot is provided on the left for a 12th magnitude G dwarf collected during the {\it{Kepler}} mission
(note the scale change by $\sim$4x).
}
\end{figure}

\begin{figure}
\includegraphics[angle=0,scale=1.0,keepaspectratio=true]{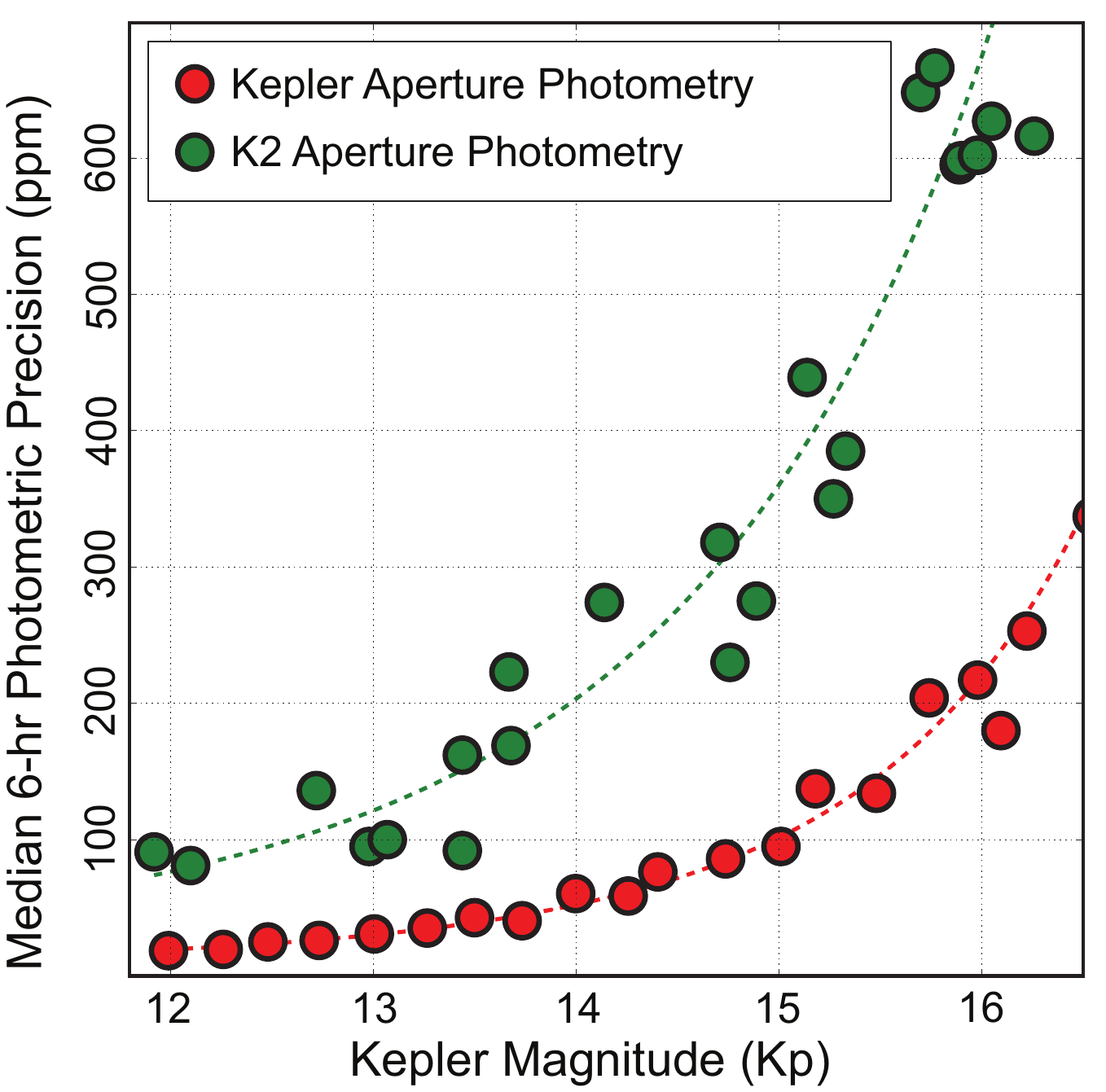}
\caption{
The December 2013 on-orbit K2 test
demonstrated a photometric precision within
a factor 3-4 times that of {\it{Kepler}}. The plot above
provides the {\it{Kepler}} and K2 median 1$\sigma$ sensitivities to
6-hour transits ($\zeta$) as a function of target magnitude.
}
\end{figure}

\begin{figure}
\includegraphics[angle=0,scale=1.0,keepaspectratio=true]{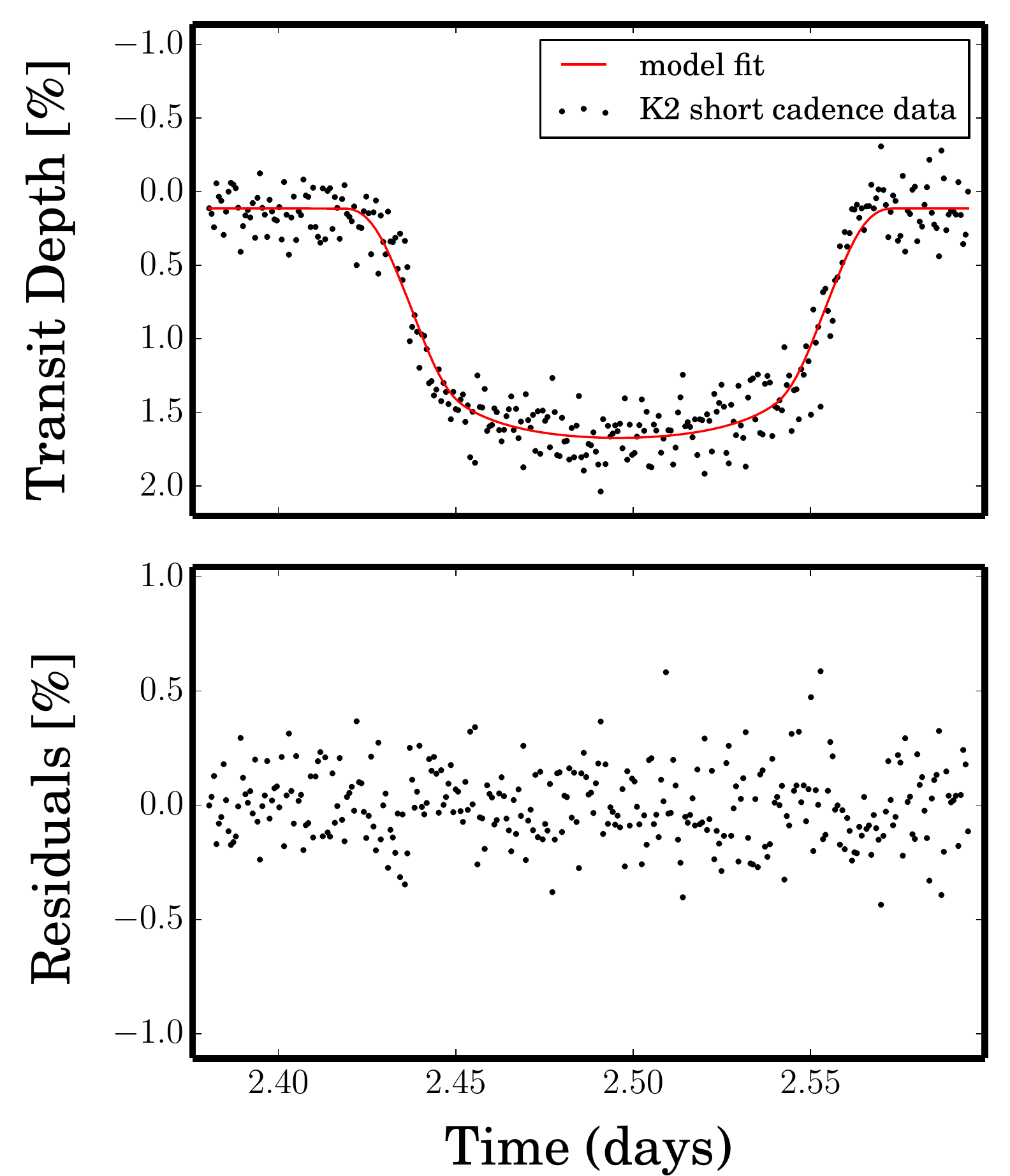}
\caption{
K2 light curve of the transiting exoplanet WASP-28b. 
During a 2.6-day science verification period, contained within our January 2014 engineering test,
we obtained 1-min sampled observations of the 12th magnitude target star. K2 observations are
shown in black and the best fitting transit model in in red.
The residuals (bottom panel) of the transit fit to the data show a point-to-point (rms) 
scatter, for a 6-hr integrated noise level, of 84 ppm. 
}
\end{figure}

\end{document}